\documentstyle[preprint,tighten,aps]{revtex} 

\def\lqcd{\Lambda_{\rm QCD}}

\begin{document}

\preprint{\vbox{ \hbox{UCSD/PTH 98--31} \hbox{hep-ph/9809423} }}

\title{$B$ decay and the $\Upsilon$ mass}

\author{Andre H.\ Hoang\footnote{ahoang@ucsd.edu \\ Address after 
  September 1998: Theory Division, CERN, CH-1211 Geneva 23, Switzerland.}, 
  Zoltan Ligeti\footnote{zligeti@ucsd.edu \\ Address after 
  October 1998: Theory Group, Fermilab, P.O.Box 500, Batavia, IL 60510.} and 
  Aneesh V.\ Manohar\footnote{amanohar@ucsd.edu \\ Address after 
  November 1998: still UCSD.} }

\address{Department of Physics, University of California at San Diego,\\
9500 Gilman Drive, La Jolla, CA 92093--0319}

\date{September 1998}

\maketitle

\begin{abstract}%
Theoretical predictions for inclusive semileptonic $B$ decay rates are
rewritten in terms of the $\Upsilon(1S)$ meson mass instead of the $b$ quark
mass, using a modified perturbation expansion.  This method gives theoretically
consistent and phenomenologically useful results.  Perturbation theory is well
behaved, and the largest theoretical error in the predictions coming from the
uncertainty in the quark mass is eliminated.  The results are applied to the
determination of $|V_{cb}|$, $|V_{ub}|$, and $\lambda_1$.

\end{abstract}

\newpage

Inclusive decay rates of hadrons containing a heavy quark can be systematically
expanded in powers of $\alpha_s(m_Q)$ and $\lqcd/m_Q$, where $m_Q$ is the mass
of the heavy quark and $\lqcd$ is the nonperturbative scale parameter of the
strong interactions. In the $m_Q\to\infty$ limit, inclusive decay rates are
given by free quark decay and the order $\lqcd/m_Q$ corrections
vanish~\cite{CGG}. The leading nonperturbative corrections of order
$\lqcd^2/m_Q^2$ are parameterized by two hadronic matrix
elements~\cite{incl,MaWi,Blok}. These results are now used to determine the CKM
matrix elements $|V_{cb}|$ and $|V_{ub}|$, using experimental data on inclusive
semileptonic $B$ meson decays.

At present, the largest theoretical uncertainties in the $B\to X_c e\bar\nu$
and $B\to X_u e\bar\nu$ decay rates arise from poor knowledge of the $b$ quark
mass. The $b$ quark pole mass is an infrared sensitive quantity which is not
well defined beyond perturbation theory~\cite{renorm}.  This is related to the
bad behavior of perturbative corrections to the inclusive decay rate when it is
written in terms of the pole mass~\cite{LSW,BBB}.  The decay rate has been
rewritten, with the hope of reducing the theoretical uncertainties, in terms of
other quantities such as the $B$ meson mass and the $\bar\Lambda$ parameter of
HQET, or in terms of the infrared safe $\overline{\rm MS}$ mass of the $b$
quark.  Nonetheless, the uncertainties remain sizable and are a significant
part of the present theoretical errors on $|V_{cb}|$ and $|V_{ub}|$.  

In this paper, the theoretical predictions for semileptonic $B$ decay rates are
rewritten in terms of the $\Upsilon(1S)$ meson mass rather than the $b$ quark
mass. This eliminates the uncertainty due to the $m_b^5$ factor in the decay
rates, and at the same time improves the behavior of the perturbation series. 
Our formulae relate measurable quantities to one another and the resulting
perturbation series is free of renormalon ambiguities.  

The inclusive decay rate $B\to X_u e\bar\nu$ is~\cite{LSW,BBB}
\begin{eqnarray}\label{bupole}
\Gamma(B\to X_u e\bar\nu) = {G_F^2 |V_{ub}|^2\over 192\pi^3}\, m_b^5\,
  \bigg[ 1 &-& 2.41 {\alpha_s\over\pi}\, \epsilon 
  - 3.22 {\alpha_s^2\over\pi^2} \beta_0\, \epsilon^2 
  - 5.18 {\alpha_s^3\over\pi^3} \beta_0^2\, \epsilon^3 - \ldots \nonumber\\
&-& {9\lambda_2 - \lambda_1 \over 2m_b^2} + \ldots \bigg] . 
\end{eqnarray}
Here $m_b$ is the $b$ quark pole mass, $\beta_0=11-2n_f/3$ is the first
coefficient of the QCD $\beta$-function, and $\alpha_s$ is the running coupling
constant in the $\overline{\rm MS}$ scheme at the scale $\mu=m_b$. The variable
$\epsilon=1$ denotes the order in our modified expansion. There is a subtlety
in the power counting for the $\Upsilon$ mass, for which the difference between
powers of $\alpha_s$ and $\epsilon$ will be important. Only the part of the
$\alpha_s^{2,3}$ corrections proportional to $\beta_0^{1,2}$ (the BLM
piece~\cite{BLM}) is known.  It is the dominant part of the two-loop correction
in examples where the entire two-loop correction has been computed (see, e.g.,
Eq.~(\ref{polemass})).  The $1/m_b^2$ terms are a few percent, so the
$\alpha_s/m_b^2$ and $1/m_b^3$ corrections are negligible.  With
$\alpha_s(m_b)=0.22$ and $n_f=4$, the perturbative series in Eq.~(\ref{bupole})
is $1-0.17\epsilon-0.13_{\rm BLM}\epsilon^2-0.12_{\rm BLM}\epsilon^3-\ldots$,
where we have used the subscript BLM to remind the reader that only the BLM
piece of the $\alpha_s^{2,3}$ terms has been computed. It is difficult to
estimate $\Gamma(B\to X_u e\bar\nu)$ reliably, since uncertainties in $m_b^5$
and in the perturbative expansion are large.  Moreover, the perturbation series
at large orders contains a contribution of order $\alpha_s^n\beta_0^{n-1}n!$
which is not Borel summable, leading to a renormalon ambiguity.  

The pole mass $m_b$ is an infrared sensitive quantity.  It can be related
to an infrared safe mass such as the $\overline{\rm MS}$ mass
$\overline{m}_b$ via (for $n_f=4$)~\cite{mass}
\begin{equation}\label{polemass}
{m_b \over \overline{m}_b(m_b)} = 1 + {4\alpha_s\over3\pi}\, \epsilon
  + (1.56\beta_0 - 1.07) {\alpha_s^2\over\pi^2}\, \epsilon^2 + \ldots 
\end{equation}
This relation also has terms of the form $\alpha_s^n\beta_0^{n-1}n!$ at high
orders. There is a cancellation between the $\alpha_s^n\beta_0^{n-1}n!$ terms
in Eqs.~(\ref{bupole}) and (\ref{polemass}) when the inclusive decay rate is
rewritten in terms of the $\overline{\rm MS}$ mass~\cite{rencan}.  While this
cancellation is present at high orders, the perturbation series in
Eq.~(\ref{bupole}) with $m_b \to \overline m_b$ is $1 +0.30\epsilon
+0.19_{\rm BLM}\epsilon^2 +0.05_{\rm BLM}\epsilon^3$~\cite{BBB}, so there are
still large corrections at low orders. Furthermore, using the $\overline{\rm
MS}$ mass does not remove the quark mass uncertainty in the decay rate. 

A simple method of avoiding problems with the quark mass is to use instead the
hadron mass. Unfortunately, the $B$ meson and $b$ quark masses differ by order
$\lqcd$, and so this reintroduces a $\lqcd/m_b$ correction to the inclusive
decay rate. A better method is to rewrite expressions like Eq.~(\ref{bupole})
in terms of the $\Upsilon$ mass to obtain well defined formulae for $B$ decay
rates in terms of $m_\Upsilon$. The resulting expressions are free of
renormalon ambiguities, and they express one measurable quantity in terms of
another. We will also see numerically that the $\alpha_s$ corrections are small
when the $B$ decay rate is written in terms of the $\Upsilon$ mass.  

There is an interesting theoretical subtlety in
the behavior of the perturbation series for the $\Upsilon$ mass in terms of the
quark pole mass. This is simplest to illustrate in the large $\beta_0$ (i.e.,
bubble summation) approximation.  Schematically, the perturbative expansion 
of the $\Upsilon$ mass in terms of $m_b$ is
\begin{eqnarray}\label{schematic}
{m_\Upsilon \over 2m_b} \sim 1 - {(\alpha_s C_F)^2\over8}\, \bigg\{ 1 
  &+& {\alpha_s\beta_0\over\pi}\, \Big( \ell + 1 \Big) 
  + \bigg({\alpha_s\beta_0\over\pi}\bigg)^2\,
  \Big( \ell^2 + \ell + 1 \Big) + \ldots \nonumber\\*
&+& \bigg({\alpha_s\beta_0\over\pi}\bigg)^n\,
  \Big( \ell^n + \ell^{n-1} + \ldots + 1 \Big) + \ldots \bigg\} \,,
\end{eqnarray}
where $\ell=\ln[\mu/(m_b\alpha_s C_F)]$, $C_F=4/3$, and the precise
coefficients are not shown.  At low orders this series is of the form
$\{\alpha_s^2,\ \alpha_s^3\beta_0,\ \alpha_s^4\beta_0^2,\ \ldots\}$, whereas
the corrections in Eqs.~(\ref{bupole}) and (\ref{polemass}) are of order
$\{\alpha_s,\ \alpha_s^2\beta_0,\ \alpha_s^3\beta_0^2,\ \ldots\}$.  An explicit
calculation using the Borel transform of the static quark potential~\cite{ugo}
shows that this mismatch disappears at higher orders.  The terms in
Eq.~(\ref{schematic}) of the form $(\ell^n+ \ell^{n-1}+ \ldots+1)$ exponentiate
to give $\exp(\ell) = \mu/(m_b\alpha_s C_F)$, and correct the mismatch between
the powers of $\alpha_s$ and $\beta_0$.  This has to happen since $m_\Upsilon$
is a physical quantity, so the renormalon ambiguities must cancel in
Eq.~(\ref{schematic}) between $2m_b$ and the potential plus kinetic
energies~\cite{andre}.

The expression for the $\Upsilon$ mass in terms of $m_b$ 
is~\cite{Upsmass},
\begin{eqnarray}\label{upsmass}
{m_\Upsilon \over 2m_b} = 1 - {(\alpha_s C_F)^2\over8} \bigg\{ 1 \epsilon 
&+& {\alpha_s\over\pi} \bigg[\bigg( \ell + \frac{11}6\bigg) \beta_0 
  - 4 \bigg] \epsilon^2  \\
&+& \bigg({\alpha_s\beta_0\over2\pi}\bigg)^2 
  \bigg( 3\ell^2 +9\ell +2\zeta(3)+\frac{\pi^2}6+\frac{77}{12}\bigg) 
  \epsilon^3 +\ldots \bigg\} . \nonumber
\end{eqnarray}
The ellipsis denote terms of order $\alpha_s^4$ with at most one power of
$\beta_0$ or $\beta_1$ (which are known), as well as terms of order
$\alpha_s^5$. The arguments following Eq.~(\ref{schematic}) show that to ensure
the cancellation of renormalon ambiguities when we combine Eqs.~(\ref{bupole})
and (\ref{upsmass}), terms of order $\alpha_s^n$ in Eq.~(\ref{upsmass}) should
be viewed as if they were only of order $\alpha_s^{n-1}$.  For this reason, the
power of $\epsilon$ in Eq.~(\ref{upsmass}) is one less than the power of
$\alpha_s$.  One should also choose the same renormalization scale, $\mu$, in
Eqs.~(\ref{bupole}) and (\ref{upsmass}).  With this prescription, it is also
expected that the infrared sensitivity present separately in
Eqs.~(\ref{bupole}) and (\ref{upsmass}) will cancel to all orders in
perturbation theory in $\epsilon$. For $\mu$ of order $m_b$,
Eq.~(\ref{upsmass}) shows no sign of convergence; for $\mu=m_b$ it yields
$m_\Upsilon = 2m_b (1 - 0.011\epsilon - 0.016\epsilon^2 - 0.024_{\rm
BLM}\epsilon^3 - \ldots)$.  The bad behavior of this series is unimportant,
since the only physical question is what happens when we use
Eq.~(\ref{upsmass}) to predict $B$ decay rates in terms of $m_\Upsilon$.

An important theoretical uncertainty in applying the above approach is the size
of nonperturbative corrections to Eq.~(\ref{upsmass}).  The dynamics of the
$\Upsilon$ system can be described using NRQCD~\cite{NRQCD}.  The leading
nonperturbative corrections to $m_\Upsilon$ arise from matrix elements in the
$\Upsilon$ of $H_{\rm light}$, the Hamiltonian of the light degrees of freedom.
In $B$ mesons, the leading nonperturbative correction to the $B$ meson mass is
due to the matrix element of $H_{\rm light}$, which is the $\bar\Lambda$
parameter of order $\lqcd$. The $\lqcd$ dependence is different for the
$\Upsilon$. $H_{\rm light}$ is the integral of a local Hamiltonian density,
$H_{\rm light}=\int\! {\rm d}^3x\, {\mathcal H}_{\rm light}(x)$. The radius of
the $\Upsilon$ is $a\sim1/(m_b \alpha_s)$, so the matrix element of $H_{\rm
light}$ is of order $a^3\lqcd^4$, by dimensional analysis. (Note that the
matrix element of ${\cal H}_{\rm light}$ is of order $\lqcd^4$, not $m_b^4$. 
Terms that grow with $m_b$ can be treated using NRQCD  perturbation theory.)
Using $1/a\sim 1\,$GeV, and $\lqcd \sim 350\,$MeV, of order a constituent quark
mass, gives a nonperturbative correction of $15\,$MeV. Using instead $\lqcd\sim
500\,$MeV gives a correction of $60\,$MeV. We will use $100\,$MeV as a
conservative estimate of the nonperturbative contribution to $m_\Upsilon$.

Substituting Eq.~(\ref{upsmass}) into Eq.~(\ref{bupole}) and collecting terms
of a given order in $\epsilon$ gives the $B\to X_u e\bar\nu$ decay rate in the
large $\beta_0$ approximation in terms of the $\Upsilon$ mass, 
\begin{eqnarray}\label{buups}
\Gamma(B\to X_u e\bar\nu) = {G_F^2 |V_{ub}|^2\over 192\pi^3}
  \bigg({m_\Upsilon\over2}\bigg)^5\, \bigg[ 1 &-& 0.115\epsilon 
  - 0.035_{\rm BLM} \epsilon^2 - 0.005_{\rm BLM} \epsilon^3 \nonumber\\
&-& {9\lambda_2 - \lambda_1 \over 2(m_\Upsilon/2)^2} + \ldots \bigg] ,
\end{eqnarray}
using $\mu=m_b$ and $\alpha_s(m_b)=0.22$.  The non-BLM parts of the
$\epsilon^{2,3}$ terms have been neglected.  The perturbation series, $1 -
0.115\epsilon - 0.035_{\rm BLM}\epsilon^2 - 0.005_{\rm BLM}\epsilon^3$, is far
better behaved than the series in Eq.~(\ref{bupole}), $1 - 0.17\epsilon -
0.13_{\rm BLM}\epsilon^2 - 0.12_{\rm BLM}\epsilon^3$, or the series expressed
in terms of the $\overline{\rm MS}$ mass, $1+0.30\epsilon+0.19_{\rm
BLM}\epsilon^2+0.05_{\rm BLM}\epsilon^3$.  The uncertainty in the $B$ decay
rate using Eq.~(\ref{buups}) is much smaller than that in Eq.~(\ref{bupole}),
both because the perturbation series is better behaved, and because the
$\Upsilon$ mass is better known (and better defined) than the $b$ quark mass.

The non-BLM order $\alpha_s^2$ corrections to $b$ decay have only been
calculated for $b\to c$ decay, at three values of the invariant mass of the
lepton pair~\cite{nonBLM}.  Extrapolating these results to $m_c\to0$ gives the
estimate that the complete $\alpha_s^2$ correction to $b\to u$ decay is about
$(90\pm10)$\% of the order $\alpha_s^2\beta_0$ result~\cite{LSW}.  With this
estimate, and including the entire $\epsilon^2$ term in Eq.~(\ref{upsmass})
gives at order $\epsilon^2$
\begin{eqnarray}\label{buups2}
\Gamma(B\to X_u e\bar\nu) = {G_F^2 |V_{ub}|^2\over 192\pi^3}
  \bigg({m_\Upsilon\over2}\bigg)^5\,
  \Big[ 1 &-& 0.115 \epsilon - (0.045\pm0.013)\epsilon^2 \nonumber\\
&-& (0.20\lambda_2 - 0.02\lambda_1)/{\rm GeV}^2 \Big] \,,
\end{eqnarray}
where the error on the $\epsilon^2$ term is due to the $\pm10\%$ uncertainty in
the $\alpha_s^2$ term in $b\to u$ decay.  Eq.~(\ref{buups2}) yields a relation
between $|V_{ub}|$ and the total semileptonic $B\to X_u e\bar\nu$ decay rate
with very small uncertainty,
\begin{eqnarray}\label{Vub}
|V_{ub}| &=& (3.06 \pm 0.08 \pm 0.08) \times 10^{-3} \nonumber\\
&& \times \left( {{\cal B}(B\to X_u e\bar\nu)\over 0.001}
  {1.6\,{\rm ps}\over\tau_B} \right)^{1/2} ,
\end{eqnarray}
where we have used $\lambda_2 = 0.12\,{\rm GeV}^2$ and $\lambda_1 =
(-0.25\pm0.25)\,{\rm GeV}^2$. The first error is obtained by assigning an
uncertainty in Eq.~(\ref{buups2}) equal to the value of the $\epsilon^2$ term
and the second is from assuming a $100\,$MeV uncertainty in
Eq.~(\ref{upsmass}).  The scale dependence of $|V_{ub}|$ due to varying $\mu$
in the range $m_b/2< \mu <2m_b$ is less than 1\%.  The uncertainty in
$\lambda_1$ makes a negligible contribution to the total error.  It is unlikely
that ${\cal B}(B\to X_u e\bar\nu)$ will be measured without significant
experimental cuts, for example, on the hadronic invariant mass~\cite{FLW}.  Our
method should reduce the uncertainties in such analyses as well.

The $B\to X_c e\bar\nu$ decay depends on both $m_b$ and $m_c$.  It is
convenient to express the decay rate in terms of $m_\Upsilon$ and $\lambda_1$
instead of $m_b$ and $m_c$, using Eq.~(\ref{upsmass}) and
\begin{equation}\label{mbmc}
m_b - m_c = \overline{m}_B - \overline{m}_D + \bigg( 
  {\lambda_1\over 2\overline{m}_B} - {\lambda_1\over 2\overline{m}_D} \bigg) 
  + \ldots \,,
\end{equation}
where $\overline{m}_B = (3m_{B^*}+m_B)/4=5.313\,$GeV and $\overline{m}_D =
(3m_{D^*}+m_D)/4=1.973\,$GeV.  The $\alpha_s$ correction to free quark decay is
known analytically~\cite{Yossi}, and the full order $\alpha_s^2$
result~\cite{nonBLM} can be estimated numerically (at the scale $\mu=m_b$) by
multiplying the order $\alpha_s^2\beta_0$ correction~\cite{LSW} by
$0.9\pm0.05$.  We then find
\begin{eqnarray}\label{bcups}
\Gamma(B\to X_c e\bar\nu) = {G_F^2 |V_{cb}|^2\over 192\pi^3}
  \bigg({m_\Upsilon\over2}\bigg)^5\, 0.533 \Big[ 1 &-& 
  0.096\epsilon - 0.031\epsilon^2 \nonumber\\
&-& (0.28\lambda_2 + 0.12\lambda_1)/{\rm GeV}^2 \Big] \,, 
\end{eqnarray}
where the phase space has also been expanded in $\epsilon$.  For comparison, 
the perturbation series in this relation when written in terms of the pole 
mass is $ 1- 0.12\epsilon- 0.06\epsilon^2 -\ldots$~\cite{LSW}. 
Equation~(\ref{bcups}) implies
\begin{eqnarray}\label{Vcb}
|V_{cb}| &=& (41.6 \pm 0.8 \pm 0.7 \pm 0.5) \times 10^{-3} \nonumber\\
&& \times \eta_{\rm QED} \left( {{\cal B}(B\to X_c e\bar\nu)\over0.105}\,
  {1.6\,{\rm ps}\over\tau_B}\right)^{1/2} ,
\end{eqnarray}
where $\eta_{\rm QED}\sim1.007$ is the electromagnetic radiative correction.
The uncertainties come from assuming an error in Eq.~(\ref{bcups}) equal to the
$\epsilon^2$ term, the $0.25\,{\rm GeV}^2$ error in $\lambda_1$, and a
$100\,$MeV error in Eq.~(\ref{upsmass}), respectively.  The second uncertainty
is reduced to $\pm0.3$ by extracting $\lambda_1$ from the electron spectrum in
$B\to X_c e\bar\nu$; see Eq.~(\ref{lambda1}).  The agreement of $|V_{cb}|$ with
other determinations (such as exclusive decays) is a check that nonperturbative
corrections to Eq.~(\ref{upsmass}) are indeed small.  

In Ref.~\cite{gremmetal} $\bar\Lambda$ and $\lambda_1$ were extracted from the
lepton spectrum in $B\to X_c e\bar\nu$ decay.  With our approach, there is no
dependence on $\bar\Lambda$, so we can determine $\lambda_1$ directly with
small uncertainty.  Considering the observable $R_1 = \int_{1.5{\rm GeV}}
E_\ell ({\rm d}\Gamma / {\rm d}E_\ell) {\rm d}E_\ell \big/ \int_{1.5{\rm GeV}}
({\rm d}\Gamma / {\rm d}E_\ell) {\rm d}E_\ell$, a fit to the same data yields
\begin{equation}\label{lambda1}
  \lambda_1 = (-0.27 \pm 0.10 \pm 0.04) \,{\rm GeV}^2 .
\end{equation}
The central value includes corrections of order $\alpha_s^2
\beta_0$~\cite{gremmetal2}.  The first error is dominated by $1/m_b^3$
corrections~\cite{gremmetal3}.  We varied the dimension-six matrix elements
between $\pm(0.5\,{\rm GeV})^3$, and combined their coefficients in quadrature
in the error estimate.  The second error is from assuming a $100\,$MeV
uncertainty in Eq.~(\ref{upsmass}).  The central value of $\lambda_1$ at tree
level or at order $\alpha_s$ is within $0.03\,{\rm GeV}^2$ of the one in
Eq.~(\ref{lambda1}).  

The above results can also be applied to $D\to X e\nu$ decay, using
$\alpha_s(m_c)=0.35$ and $n_f=3$. Nonperturbative effects are clearly much
larger in the $J/\psi$ than in the $\Upsilon$, so one might expect the entire
analysis to break down completely.  It is remarkable that this does not occur. 
Using $m_{J/\psi} = 2m_c (1 - 0.027\epsilon - 0.059\epsilon^2 - 0.130\epsilon^3
- \ldots)$, neglecting $m_s$, and following the same procedure as for $b\to u$
decay, we find
\begin{eqnarray}\label{charm2}
\Gamma(D\to X e\nu) = {G_F^2 (|V_{cs}|^2+|V_{cd}|^2)\over 192\pi^3} 
  \bigg({m_{J/\psi}\over2}\bigg)^5 \Big[ 1 &-& 
  0.13 \epsilon - 0.03 \epsilon^2 \nonumber\\
&-& (1.9\lambda_2 - 0.2\lambda_1)/{\rm GeV}^2 \Big] . 
\end{eqnarray}
The $\epsilon^3$ contribution to Eq.~(\ref{charm2}) is larger than the
order $\epsilon^2$ term.  The perturbation series expressed in terms of the
pole mass has a much worse behavior, roughly $1-0.27\epsilon-0.32\epsilon^2$. 
Using $\lambda_2(m_c) = 0.14\,{\rm GeV}^2$ and $\lambda_1$ from
Eq.~(\ref{lambda1}), we obtain
\begin{eqnarray}\label{Vcs}
|V_{cs}|^2 + |V_{cd}|^2 &=& (1.00 \pm 0.06 \pm 0.04) \nonumber\\
&& \times \left( {{\cal B}(D^\pm\to X e\nu)\over 0.17}\,
  {1.06\,{\rm ps}\over\tau_{D^\pm}}\right) ,
\end{eqnarray}
where the uncertainties come from assuming an error in Eq.~(\ref{charm2}) equal
to the $\epsilon^2$ term and the error in $\lambda_1$, respectively.  We have
not included an estimate of nonperturbative corrections to the $J/\psi$ mass,
or of scale dependence.  The LEP measurements of the hadronic $W$ width yield
$|V_{cs}| = 0.98\pm 0.07\pm 0.04$~\cite{Wwidth}.  The uncertainty in
Eq.~(\ref{Vcs}) is comparable to this, since the experimental error of ${\cal
B}(D^\pm\to X e\nu)$ is about 10\%.  Eq.~(\ref{Vcs}) has theoretical
uncertainties which we cannot estimate. The validity of quark-hadron duality
may be questionable since the final states are almost saturated by $K$ and
$K^*$. In addition, an estimate similar to that for the $\Upsilon$ suggests
that the nonperturbative contribution to the $J/\psi$ mass is of order
$500\,$MeV (using $1/a\sim0.5\,$GeV and $\lqcd\sim500\,$MeV). This gives an
uncertainty of order 100\% in $|V_{cs}|^2 + |V_{cd}|^2$. The agreement of
Eq.~(\ref{Vcs}) with the experimental results may be a coincidence, or may
signal that nonperturbative corrections in the mass relation are much smaller
than naive expectations.

We have chosen to write our $B$ decay results in terms of the $\Upsilon(1S)$
mass. One could equally well write them in terms of the mass of excited states,
such as the $\Upsilon(2S)$. The perturbation series is expected to be worse
behaved than for the $\Upsilon(1S)$. The main difference is in the estimate of
nonperturbative corrections to the $\Upsilon(2S)$ mass. The radius of the $2S$
state is about four times that of the $1S$, so the nonperturbative corrections,
which grow as $a^3$, are approximately 64 times larger. This implies a similar
increase in the error on the CKM angles. Ignoring nonperturbative corrections
for the moment, the analog of Eq.~(\ref{upsmass}) for the $\Upsilon(2S)$
evaluated at the scale $\mu=m_b$ is $m_{\Upsilon(2S)} = 2m_b(1 -
0.0027\epsilon - 0.0059\epsilon^2 - 0.0117_{\rm BLM}\epsilon^3- \ldots)$.
Numerically, the first few corrections are smaller than for the $\Upsilon(1S)$,
but the convergence of the series is worse.  The $B\to X_u e\bar\nu$ decay rate
in the large $\beta_0$ approximation in terms of the $\Upsilon(2S)$ mass is then
\begin{eqnarray}\label{bu2ups}
\Gamma(B\to X_u e\bar\nu) &=& {G_F^2 |V_{ub}|^2\over 192\pi^3}
  \bigg({m_{\Upsilon(2S)}\over2}\bigg)^5 \nonumber\\
&& \times \Big[ 1 - 0.155 \epsilon - 0.098_{\rm BLM} \epsilon^2
  -0.065_{\rm BLM} \epsilon^3 - \ldots \Big] . 
\end{eqnarray}
Compared to Eq.~(\ref{buups}), the convergence is worse, as expected. 
Nevertheless, even this formula gives a reasonable extraction of $|V_{ub}|$. 
The ratio of $|V_{ub}|^2$ extracted using the $2S$ and $1S$ masses is ${\rm
[Eq.~(\ref{bu2ups})] / [Eq.~(\ref{buups})]} =\{1.34,\ 1.27,\ 1.17,\ 1.08\}$,
where the $n$th number is obtained by truncating both equations at order
$\epsilon^{n-1}$, and neglecting the $\lambda_{1,2}$ corrections. The large
difference at ``tree level", $(m_{\Upsilon(2S)} / m_\Upsilon)^5 = 1.34$, is
reduced by the series of perturbative corrections.  Expressing the $B\to X_c
e\bar\nu$ decay rate in terms of the $\Upsilon(2S)$ mass, the perturbative
corrections in Eq.~(\ref{bcups}) become
\begin{eqnarray}\label{bc2ups}
\Gamma(B\to X_c e\bar\nu) &=& {G_F^2 |V_{cb}|^2\over 192\pi^3}
  \bigg({m_{\Upsilon(2S)}\over2}\bigg)^5 \nonumber\\
&& \times 0.447\, \Big[ 1 - 0.107 \epsilon - 0.046 \epsilon ^2 + 
  \ldots \Big] . 
\end{eqnarray}
Again, the convergence of the series becomes worse.  However, the ratio of
$|V_{cb}|^2$ extracted using the $2S$ and $1S$ masses is consistent with
our estimates of the uncertainties, ${\rm [Eq.~(\ref{bc2ups})] /
[Eq.~(\ref{bcups})]} = \{1.12,\ 1.10,\ 1.08\}$, where the $n$th number is
obtained by truncating both expressions at order $\epsilon^{n-1}$.  The
difference between the $\Upsilon(2S)$ and $\Upsilon(1S)$ results provide an
estimate of nonperturbative contributions to the $\Upsilon$ mass. They suggest
that nonperturbative effects are smaller than the conservative estimate we have
used; they are certainly much smaller than the naive estimate above of a
$64\times 100\,{\rm MeV}=6.4\,$GeV nonperturbative contribution to the
$\Upsilon(2S)$ mass.

We have shown that inclusive semileptonic $B$ decay rates can be predicted in
terms of the $\Upsilon(1S)$ mass instead of the $b$ quark mass.  It is crucial
to our analysis to use the modified expansion in $\epsilon$ rather than the
conventional expansion in powers of $\alpha_s$.  Our formulae relate only
physical quantities to one another. They result in smaller theoretical
uncertainties than existing numerical predictions, and the behavior of the
perturbation series is improved.  Moreover, the uncertainties can be estimated
without resorting to cumbersome arguments, and they can be checked using the
experimental data.  

Our main results are Eqs.~(\ref{Vcb}) and (\ref{Vub}), which relate the total
semileptonic $B\to X_{c,u} e\bar\nu$ decay rates to $|V_{cb}|$ and $|V_{ub}|$. 
The uncertainties are below 5\% at present, and it may be possible to reduce
them further.  Our determination of $\lambda_1$ is given in
Eq.~(\ref{lambda1}).  We hope that applications of the method introduced in
this paper will prove useful --- besides reducing the uncertainties of
$|V_{cb}|$ and $|V_{ub}|$ --- in analyzing a large class of data emerging from
present and future $B$ decay experiments.  Details of our method, as well as
other applications, such as to nonleptonic and exclusive semileptonic $B$
decays, will be discussed elsewhere~\cite{next}.

We thank Mark Wise for useful discussions.  
This work was supported in part by the DOE grant DOE-FG03-97ER40546 
and by the NSF grant PHY-9457911.

\end{document}